# A Close Look at the EPR Data of Weihs et al

James H. Bigelow
jlandjhb@gmail.com
June 27, 2009

**ABSTRACT**

I examine data from EPR experiments conducted in 1997 through 1999 by Gregor Weihs and colleagues. They used detection windows of 4-6 ns to identify coincidences; I find that one obtains better results with windows 40-50 ns wide. Coincidences identified using different windows have substantially different distributions over the sixteen combinations of Alice's and Bob's measurement settings and results, which is the essence of the coincidence time loophole. However, wide and narrow window coincidences violate a Bell inequality equally strongly. The wider window yields substantially smaller violations of no-signaling conditions.

**ACKNOWLEDGEMENT**

In this paper I examine data from EPR experiments conducted in 1997 through 1999 by Gregor Weihs, Thomas Jennewein, Christoph Simon, Harald Weinfurter, and Anton Zeilinger. Weihs and his colleagues have made these data freely available to me, as they have previously to others (e.g., Adenier & Khrennikov, 2007; Zhao et al, 2007). I am grateful to them and I heartily recommend that other experimenters adopt the same practice.

**INTRODUCTION**

In the most commonly performed EPR experiment—an experiment that tests a Bell (1964) inequality—a source generates pairs of photons. Each photon travels to one of two observers Alice and Bob, who each perform one of two measurements and obtain one of two results. The data analysis in these experiments generally consists of only two steps: identifying pairs of detections (called coincidences), one from each observer, that correspond to the pairs of photons generated by the source; and calculating the correlations between the results obtained by the observers. These correlations are then inserted into one of the many Bell-type inequalities to determine whether or not the inequality is violated.

From my first reading about EPR experiments I have felt that more information could be extracted from these experiments. This is not possible when the task of identifying coincidences is automated by a coincidence circuit (e.g., Aspect, 2002), and the identified coincidences are the only recorded data. But Weihs and his colleagues registered the individual detections on both sides completely independently (Wiehs et al, 1998), and this opens the way for a more complete analysis. I have used their data (hereafter called the WJSWZ data, from the initials of the authors of the 1998 paper), to test alternative rules for identifying coincidences. I also estimate the number of identified coincidences that are not true (false positives), though not the number of true coincidences that are not identified (false negatives). I have also been able to shed some light on the so-called coincidence time loophole (e.g., see De Raedt et al, 2007; Larsson & Gill, 2004; Morgan, 2008; Willebordse, 2008; Zhao et al, 2007). According to this hypothesis, the time difference between the two detections of a true coincidence might be correlated with the measurements performed and/or the results obtained on the two sides.

## THE DATA

The WJSWZ experiments are described in Weihs et al (1998) and in more detail in Weihs (2007). A source generated pairs of photons, each of which traveled via an optical fiber to one of two observers Alice and Bob. There it first passed through an electro-optical modulator, then a polarizer, and finally produced a "click" in one of two detectors. Applying a voltage rotated the polarization of light passing through the modulator. Two voltages were chosen modulator, giving two rotations labeled $s_A$=0 or 1 (for Alice's setting) and $s_B$=0 or 1 (for Bob's setting). The detectors were labeled $r_A$=0 or 1 (for Alice's result) and $r_B$=0 or 1 (for Bob's result). The settings $s_A$ and $s_B$ were randomly chosen anew every 100 ns. (The average time between detections on each side of the experiment was tens of thousands of ns, so a new setting was usually chosen for each detection.) Detections that occurred during the 14 ns switching periods were discarded. Over the course of the experiment, Alice and Bob each recorded the times ($t_A$ for Alice, $t_B$ for Bob), settings, and results for thousands of detections. Thus the data produced by each experiment consists of two sequences of triples:

(1a) $\quad Alog = \{(t_{A,k}, s_{A,k}, r_{A,k}), k = 1,..., N_A\}$

(1b) $\quad Blog = \{(t_{B,l}, s_{B,l}, r_{B,l}), l = 1,..., N_B\}$

The WJSWZ data include Alice's and Bob's detection logs from a substantial number of experiments. One series of 25 experiments was called "longdist," and was described as the "main run of long-distance Bell inequality tests." One of these experiments, longdist35, occurred on April 22, 1998, and collected 388,455 Alice detections and 302,271 Bob detections over a run of ten seconds in length. I give detailed results for this particular experiment because it was the WJSWZ experiment analyzed in Zhao et al (2007). I performed the same analysis on 30 other experiments, and the results for longdist35 are typical.

# RULES FOR IDENTIFYING COINCIDENCES

I consider three rules for identifying a pair of detections, one from each log, as a coinicidence. The rule I use in the remainder of this paper accepts as a coincidence any pair of detections $(t_{A,k}, s_{A,k}, r_{A,k}) \in Alog$ and $(t_{B,l}, s_{B,l}, r_{B,l}) \in Blog$ for which $u \leq t_B - t_A \leq v$ for specified $u$ and $v$. For comparison with the other two rules, it will be convenient to describe this rule in an equivalent but slightly different form. Let $\delta = (v+u)/2$ and $w = (v-u)/2$. Then this rule accepts any pair of detections that occur within a time $w$ of one another, after Alice's log has been shifted by a time $\delta$. I call this rule *allpr* (for all pairs), and it is the most inclusive rule I consider. The resulting set of coincidences is:

$$C_{allpr}(w, \delta) = \{(k,l) \mid |t_{B,l} - (t_{A,k} + \delta)| \leq w\}$$

The rule used by most authors (e.g., Weihs et al, 1998; Zhao et al, 2007) includes the subset of pairs in $C_{allpr}(w, \delta)$ for which the detections occur in sequence, without a third detection occurring in between. It yields this set of coincidences:

$$C_{inseq}(w, \delta) = \left\{(k,l) \,\middle|\, \begin{array}{l} |t_{B,l} - (t_{A,k} + \delta)| \leq w \\ (t_{A,k} + \delta) > t_{B,l-1} \\ (t_{A,k+1} + \delta) > t_{B,l} \\ (t_{A,k-1} + \delta) < t_{B,l} \\ (t_{A,k} + \delta) < t_{B,l+1} \end{array}\right\}$$

The first condition ensures that the pair $(k,l) \in C_{allpr}(w, \delta)$. If $(t_{A,k} + \delta) < t_{B,l}$, then the second and third conditions ensure that neither Bob's detection $l-1$ nor Alice's detection $k+1$ occur between Alice's detection $k$ and Bob's detection $l$. The fourth and fifth conditions are trivially satisfied. If instead $(t_{A,k} + \delta) > t_{B,l}$, then the second and third conditions are trivially satisfied, while the fourth and fifth ensure that neither Alice's detection $k-1$ nor Bob's detection $l+1$ occur between Alice's detection $k$ and Bob's detection $l$.

The first two rules permit a detection to participate in two or more identified coincidences, whereas a detection cannot be a part of more than one true coincidence. The third rule identifies only pairs $(k,l) \in C_{allpr}(w, \delta)$ for which neither Alice's detection $k$ nor Bob's detection $l$ can be paired under this rule with any other detections than each other. I call such pairs *unambiguous*. The resulting set of coincidences is:

$$C_{unamb}(w, \delta) = \left\{(k,l) \,\middle|\, \begin{array}{l} |t_{B,l} - (t_{A,k} + \delta)| \leq w \\ |t_{B,l-1} - (t_{A,k} + \delta)| > w \\ |t_{B,l+1} - (t_{A,k} + \delta)| > w \\ |t_{B,l} - (t_{A,k-1} + \delta)| > w \\ |t_{B,l} - (t_{A,k+1} + \delta)| > w \end{array}\right\}$$

The first condition ensures that $(k,l) \in C_{allpr}(w,\delta)$. The second and third conditions ensure that Alice's detection $k$ does not form a coincidence with either of Bob's detections $l-1$ or $l+1$, and this in turn implies that it cannot form a coincidence with any of Bob's earlier or later detections. Similarly, the fourth and fifth conditions ensure that Bob's detection $l$ does not form a coincidence with any of Alice's detections save detection $k$. I leave it to the reader to verify that $C_{unamb}(w,\delta) \subseteq C_{inseq}(w,\delta)$.

Table 1 shows the number of coincidences identified by each of the three rules in the longdist35 experiment, for a wide range of window sizes. For this table I have taken the synchronization parameter, $\delta$, to equal 3.8 ns, the value used by WJSWZ.

**Table 1: Coincidence Counts Using the Three Rules, for the longdist35 Experiment**

| w (ns) | Allpr | Inseq | Unamb |
|---|---|---|---|
| 1 | 11,824 | 11,824 | 11,824 |
| 2 | 14,574 | 14,573 | 14,572 |
| 3 | 15,149 | 15,148 | 15,147 |
| 4 | 15,449 | 15,448 | 15,445 |
| 5 | 15,672 | 15,671 | 15,666 |
| 10 | 16,281 | 16,277 | 16,259 |
| 20 | 17,384 | 17,372 | 17,321 |
| 50 | 18,669 | 18,630 | 18,515 |
| 100 | 19,742 | 19,677 | 19,490 |
| 200 | 21,958 | 21,828 | 21,443 |
| 500 | 29,251 | 28,867 | 27,717 |
| 1,000 | 40,799 | 39,768 | 36,693 |
| 2,000 | 63,740 | 60,732 | 51,386 |
| 5,000 | 133,258 | 117,484 | 72,894 |
| 10,000 | 250,445 | 191,194 | 68,175 |
| 20,000 | 484,723 | 279,126 | 31,291 |
| 50,000 | 1,189,710 | 353,784 | 1,027 |
| 75,000 | 1,777,749 | 362,109 | 42 |

For modest window sizes $w$ it hardly matters which rule one uses. Certainly the rules identify almost the same sets of coincidences for the window size used by WJSWZ (half-width $w$ of 2 ns), and for the larger window I will recommend later (half-width $w$ of about 23 ns).

But the three rules are quite different for very large windows. Once $w$ reaches 500 ns, the *allpr* rule adds an average of 23.48 coincidences for each ns increase in $w$. The *inseq* rule, by contrast, adds very few coincidences once $w$ becomes large. As $w$ increases, it becomes more and more likely that a third detection will occur between a pair of detections separated by a time $w$. However, a pair that passes the test for one value of $w$ will pass it for all larger values of $w$, so the number of coincidences will not decrease as $w$ increases. The *unamb* rule is even more extreme; it drops coincidences as $w$ becomes large. For large enough $w$, each of Alice's detections will be within $w$ of at least two of Bob's detections, and the rule will identify no coincidences at all.

# ESTIMATING THE FALSE-POSITIVE RATE

The set of coincidences identified by any rule will include some pairs of detections that do not correspond to photons paired at the source. Aspect (2002) describes how he determined the rate of these false-positives (he termed it the accidental rate) for the experiments he conducted for his thesis in 1980-1982:

> "Additionally, a standard coincidence circuit with a 19 ns coincidence window monitored the rate of coincidences around null delay, while a delayed-coincidence channel monitored the accidental rate."

The false positive rate obtained by Aspect's method could depend on the delay time, so I have developed a simple argument for determining what the "true" false positive rate should be.[1] One can form $388,455 \times 302,271 = 1.174 \times 10^{11}$ pairs of detections from the two detection logs of the longdist35 experiment. On an XY-plot, where X and Y are Alice's and Bob's detection times, respectively, these points will all fall in a $10^{10} \times 10^{10}$ ns square. If Alice's and Bob's detections were each distributed uniformly over the ten seconds of the experiment, and if the two logs were unrelated, then the points ought to be distributed uniformly over this square. For example, the strip of all pairs for which Bob's detection occurs between zero and 0.5 ns after Alice's detection (see Fig. 1) should contain one of every $2 \times 10^{10}$ points, or 5.87 points on average.

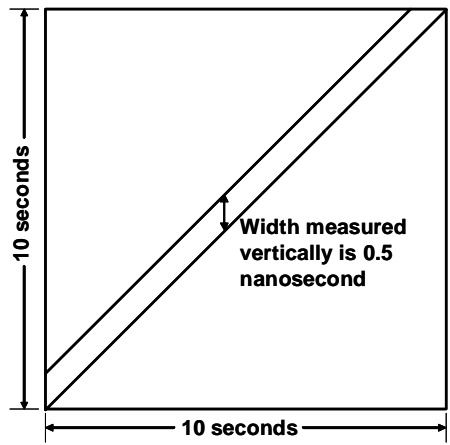

**Figure 1: Visualization of the Strip** $S(u = 0, h = 0.5)$

More generally, the strip $S(u,h) = \{(t_A, t_B) | u \leq t_B - t_A \leq u + h\}$ will contain all pairs for which Bob's detection occurs between $u$ and $u + h$ ns after Alice's detection. Even if $u$ and $h$ are as large as a millisecond (a million ns, an enormous time for the purpose at hand), the fraction of the square covered by this strip can be taken as $10^{-10} \times h$.

Of course, two strips cannot be expected to contain precisely the same number of points. A horizontal strip, for example, must contain an integer multiple of 388,455 points (the number of detections by Alice). However, my strips are diagonal and narrow enough that they don't contain a substantial portion of any horizontal (or vertical) strip. So I conjecture that the distribution of the number of points per diagonal strip should be approximately Poisson, as if they had been tossed into the square at random.

---

[1] This argument works for the *allpr* rule for all detection time differences $t_B - t_A$, but fails for the *inseq* and *unamb* rules if the detection time difference is large. It is for this reason that I elect to use the *allpr* rule.

To test this conjecture, I formed the detection pairs in all the strips of width 0.5 ns, from 75,000 ns below the diagonal (Bob's detection precedes Alice's) to 75,000 ns above the diagonal (Alice's detection precedes Bob's). For strips more than 500 ns above or below the diagonal, these statistics are entirely consistent with a Poisson distribution with mean 5.87. Moreover, the numbers of points in a strip are not correlated with the numbers of points in nearby strips. Thus it would seem that, for the longdist35 experiment, the false-positive rate is indeed about 11.74 pairs per ns of window width.[2]

## **CHOOSING A DETECTION WINDOW**

Figures 2 plot the numbers of points per strip for detection time differences for which total counts exceed estimated false positives. The left-hand chart is plotted at low enough resolution to show the entire peak, and it confirms the choice $1.8 \leq t_B - t_A \leq 5.8$ ns that WJSWZ made for a detection window for the longdist35 experiment. (WJSWZ would write the choice as $|t_B - (t_A + \delta)| \leq w$, for $\delta = 3.8$ ns and $w = 2$ ns.) But when the vertical scale is expanded (the right-hand chart), one sees that the coincidence rate exceeds the estimated false positive rate over a much wider range, say $-20.5 \leq t_B - t_A \leq 25$ ns. If we accept as coincidences all pairs that fall in this wider detection window, surely we will lose fewer true coincidences.

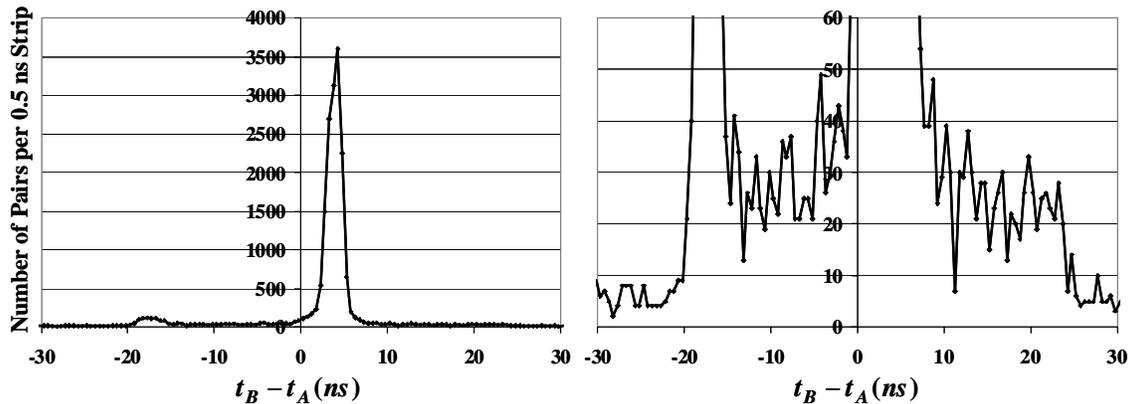

**Figure 2: Coincidence Peak for longdist35 experiment, at Low (left) and High (right) Resolution**

Tables 2 and 3 describe ranges of detection time differences in which the number of pairs significantly exceeds the estimated false positive rate. The three ranges labeled Low 1, WJSWZ Window, and High 1 comprise the wider window, $-20.5 \leq t_B - t_A \leq 25$, that I suggest as a replacement for the WJSWZ window. I include the ranges Low 2,

---

[2] As one should expect, this is half the number of coincidences that Table 1 shows are added to the *allpr* count per ns increase in $w$ — half because $w$ is the half-width of the detection window.

High 2, and High 3 for $t_B - t_A$ near ±470 ns because I find them mysterious. They contribute too few coincidences to matter for determining whether a Bell inequality is violated.

**Table 2: Ranges of Detection Time Differences, for longdist35 Experiment**

|  | Low 2 | Low 1 | WJSWZ Window | High 1 | High 2 | High 3 |
|---|---|---|---|---|---|---|
|  |  | \multicolumn{3}{c}{Wide Window $-20.5 \leq t_B - t_A \leq 25$} |  |  |  |
| Low Edge (ns) | -470.5 | -20.5 | 1.8 | 5.8 | 471.5 | 484.0 |
| High Edge (ns) | -461.5 | 1.8 | 5.8 | 25.0 | 477.5 | 487.5 |
| Observed Coincidences | 201 | 2251 | 14574 | 1223 | 174 | 141 |
| Exp False-Pos Coinc | 106 | 262 | 47 | 225 | 70 | 41 |
| z-score | 9 | 123 | 2120 | 66 | 12 | 16 |

For each range, Table 2 shows the observed number of coincidences in the range, the expected number of false positive coincidences (at a rate of 11.74 false positive coincidences per ns of window width), and how far the observed coincidence count is above the expected false positive count, measured in standard deviations of the false-positive distribution (the z-score). As noted earlier, the false-positive distribution is Poisson, so the standard deviation is the square root of the mean.

**Table 3: Raw Coincidence Counts for the Ranges of Detection Time Differences**

| $s_A$ | $s_B$ | $r_A$ | $r_B$ | Low 2 | Low 1 | WJSWZ Window | High 1 | High 2 | High 3 |
|---|---|---|---|---|---|---|---|---|---|
| 0 | 0 | 0 | 0 | 12 | 21 | 313 | 16 | 2 | 8 |
|  |  |  | 1 | 8 | 31 | 1978 | 39 | 51 | 0 |
|  |  | 1 | 0 | 17 | 68 | 1728 | 31 | 8 | 48 |
|  |  |  | 1 | 46 | 26 | 352 | 19 | 7 | 3 |
|  | 1 | 0 | 0 | 8 | 26 | 418 | 37 | 11 | 2 |
|  |  |  | 1 | 10 | 33 | 1577 | 416 | 3 | 4 |
|  |  | 1 | 0 | 47 | 84 | 1684 | 33 | 5 | 3 |
|  |  |  | 1 | 10 | 28 | 361 | 79 | 3 | 5 |
| 1 | 0 | 0 | 0 | 7 | 82 | 1636 | 25 | 4 | 20 |
|  |  |  | 1 | 4 | 30 | 294 | 19 | 50 | 4 |
|  |  | 1 | 0 | 2 | 140 | 179 | 19 | 3 | 31 |
|  |  |  | 1 | 7 | 711 | 1143 | 10 | 10 | 1 |
|  | 1 | 0 | 0 | 12 | 31 | 269 | 34 | 2 | 2 |
|  |  |  | 1 | 2 | 49 | 1386 | 354 | 5 | 2 |
|  |  | 1 | 0 | 5 | 736 | 1100 | 39 | 5 | 3 |
|  |  |  | 1 | 4 | 155 | 156 | 53 | 5 | 5 |
|  |  |  | Total | 201 | 2251 | 14574 | 1223 | 174 | 141 |

Table 3 shows the distribution of the raw coincidence counts in each range over the sixteen combinations of settings and results.[3] Note that expanding the window increases the counts by different factors in different cells, demonstrating that the coincidences identified in the two windows cannot both be fair samples. For example, the coincidences for $s_A = 0, s_B = 0, r_A = 0, r_B = 1$ increase from 1,978 to 2,048 (four percent), while the coincidences for $s_A = 1, s_B = 1, r_A = 1, r_B = 0$ increases from 1,100 to 1,875 (70 percent).

## BELL INEQUALITY VIOLATIONS

The point of an EPR experiment is to test whether, as Quantum Theory predicts, nature can violate a Bell-type inequality. The alternative is that nature is constrained to obey the principles of locality and realism so beloved of classical physics. As Table 5 shows,[4] using the wider coincidence window does not alter the fact that the WJSWZ data violate a Bell inequality. Weihs et al (1998) look for violations of the standard CHSH inequality (Clauser et al, 1969), which by permutation of the settings can be either of the following two expressions:

(2a) $\quad |E(s_A = 0, s_B = 0) - E(s_A = 1, s_B = 0)| + |E(s_A = 0, s_B = 1) + E(s_A = 1, s_B = 1)| \leq 2$

(2b) $\quad |E(s_A = 0, s_B = 1) - E(s_A = 1, s_B = 1)| + |E(s_A = 0, s_B = 0) + E(s_A = 1, s_B = 0)| \leq 2$

$E(s_A, s_B)$ is the frequency that the results agree minus the frequency they disagree, calculated over all coincidences for which Alice's setting is $s_A$ and Bob's is $s_B$. The last two lines of Table 4 give the values and standard errors of the expression on the left-hand side of the above inequalities. Inequality (2a) is violated for both the WJSWZ and Wide windows, by almost 30 standard errors.

**Table 4: Test of a Bell Inequality for longdist35 Experiment**

|  | WJSWZ Window $1.8 \leq t_B - t_A \leq 5.8$ | | Wide Window $-20.5 \leq t_B - t_A \leq 25$ | |
| --- | --- | --- | --- | --- |
|  | Value | SE | Value | SE |
| $E(s_A = 0, s_B = 0)$ | -0.69572 | 0.010865 | -0.67676 | 0.010829 |
| $E(s_A = 0, s_B = 1)$ | -0.61436 | 0.012414 | -0.60260 | 0.011548 |
| $E(s_A = 1, s_B = 0)$ | 0.70910 | 0.012365 | 0.68237 | 0.011163 |
| $E(s_A = 1, s_B = 1)$ | -0.70800 | 0.013089 | -0.67996 | 0.011102 |
| Inequality (2a) | 2.72718 | 0.024420 | 2.64169 | 0.022327 |
| Inequality (2b) | 0.10703 | 0.024420 | 0.08297 | 0.022327 |

---

[3] The counts obtained by WJSWZ are slightly different than those shown for the Narrow Window range because WJSWZ used the *inseq* rule, while I am using the *allpr* rule.

[4] Entries in Tables 4 and 5 can all be calculated from the data in Table 3. The standard error of the estimate of a probability $p$ is calculated using the formula $SE = \sqrt{p \times (1 - p) / N}$.

It is the essence of the coincidence time loophole that different detection windows identify different distributions of coincidences across the sixteen combinations of settings and results. Most authors (e.g., see De Raedt et al, 2007; Larsson & Gill, 2004; Morgan, 2008; Willebordse, 2008; Zhao et al, 2007) employ it as a mechanism by which a classical model can obtain unrepresentative samples of coincidences that violate Bell inequalities. In every WJSWZ experiment I analyzed (31 in total), I observed substantially different distributions for the narrow WJSWZ window versus a wide window, demonstrating that nature does employ this mechanism. However, substituting wide window coincidence counts for WJSWZ's narrow window coincidence counts causes very little change to the degree to which a Bell Inequality is violated. This undermines the use generally made of the coinicidence time loophole.

## NO-SIGNALING VIOLATIONS

The WJSWZ data violate the no-signaling conditions. Choosing a wider detection window reduces the violations (Table 5), but does not eliminate them entirely.

**Table 5: Test of No-Signaling Conditions for longdist35 Experiment**

|  | WJSWZ Window $1.8 \leq t_B - t_A \leq 5.8$ | | Wide Window $-20.5 \leq t_B - t_A \leq 25$ | |
|---|---|---|---|---|
|  | Value | SE | Value | SE |
| $P(r_A = 0 | s_A = 0, s_B = 0)$ | 0.524136 | 0.007554 | 0.518823 | 0.007349 |
| $P(r_A = 0 | s_A = 0, s_B = 1)$ | 0.493812 | 0.007866 | 0.524916 | 0.007226 |
| Delta | 0.030324 | 0.010906 | -0.006090 | 0.010307 |
| $P(r_A = 0 | s_A = 1, s_B = 0)$ | 0.593481 | 0.008613 | 0.486474 | 0.007633 |
| $P(r_A = 0 | s_A = 1, s_B = 1)$ | 0.568533 | 0.009180 | 0.486703 | 0.007568 |
| Delta | 0.024948 | 0.012588 | -0.000230 | 0.010749 |
| $P(r_B = 0 | s_A = 0, s_B = 0)$ | 0.466941 | 0.007546 | 0.471008 | 0.007342 |
| $P(r_B = 0 | s_A = 1, s_B = 0)$ | 0.558118 | 0.008708 | 0.485308 | 0.007632 |
| Delta | -0.091180 | 0.011523 | -0.014300 | 0.010591 |
| $P(r_B = 0 | s_A = 0, s_B = 1)$ | 0.520297 | 0.007860 | 0.477806 | 0.007228 |
| $P(r_B = 0 | s_A = 1, s_B = 1)$ | 0.470285 | 0.009251 | 0.506419 | 0.007570 |
| Delta | 0.050012 | 0.012139 | -0.028610 | 0.010466 |

While it is quite respectable to believe that nature can violate a Bell-type inequality, few, if any, physicists take seriously the notion that nature could violate a no-signaling condition. Such a violation could be used to send messages faster than light, and even to send messages to one's past self! But Adenier & Khrennikov (2007) concluded that one must abandon either the no-signaling or the fair sampling assumption. And abandoning fair sampling raises difficulties of its own.

I demonstrate that Adenier & Khrennikov were mistaken; the coincidence counts identified in either window—I'll illustrate for the wide window—can be modeled as a fair sample drawn from a set of sixteen coincidence counts that obey the no-signaling conditions. Denote by $W_{tot}(s_A, s_B, r_A, r_B)$ the number of coincidences identified in the wide window for settings and results $s_A, s_B, r_A, r_B$. Some of these coincidences correspond to pairs of photons generated by the source, which I call true coincidences and denote by $W_T(s_A, s_B, r_A, r_B)$. The remainder are false positive coincidences, which I denote by $W_F(s_A, s_B, r_A, r_B)$. Then:

(3) $\quad W_{tot}(s_A, s_B, r_A, r_B) = W_T(s_A, s_B, r_A, r_B) + W_F(s_A, s_B, r_A, r_B)$

The values of the $W_{tot}(s_A, s_B, r_A, r_B)$ can be calculated from Table 3. I estimate false positives for each combination of settings and results in the same way as I estimated total false positives, but using the singles counts for the individual settings and results (see Table 6) rather than the overall singles counts. To estimate the true coincidences $W_T(s_A, s_B, r_A, r_B)$ I subtract the expected false positive coincidences for the 45.5 ns width of the wide window from the total coincidences.

**Table 6: Singles Counts by Setting and Result**

| Alice's Detections | | Bob's Detections | |
|---|---|---|---|
| $s_A/r_A$ | # Singles | $s_B/r_B$ | # Singles |
| 0/0 | 104,122 | 0/0 | 77,988 |
| 0/1 | 100,144 | 0/1 | 74,935 |
| 1/0 | 96,348 | 1/0 | 75,892 |
| 1/1 | 90,841 | 1/1 | 73,456 |

Let $C(s_A, s_B, r_A, r_B)$ be the coincidence counts that would have been obtained if the detectors were perfect. I wish to find detection probabilities $\lambda_A(s_A, r_A)$ and $\lambda_B(s_B, r_B)$ that satisfy:

(4a) $\quad W_T(s_A, s_B, r_A, r_B) = \lambda_A(s_A, r_A) \times \lambda_B(s_B, r_B) \times C(s_A, s_B, r_A, r_B)$

or equivalently:

(4b) $\quad C(s_A, s_B, r_A, r_B) = \dfrac{W_T(s_A, s_B, r_A, r_B)}{\lambda_A(s_A, r_A) \times \lambda_B(s_B, r_B)}$

The counts $C(s_A, s_B, r_A, r_B)$ must obey the no-signaling conditions, which can be expressed as:

(5a) $\quad \dfrac{C(0,0,0,0) + C(0,0,0,1)}{C(0,0,1,0) + C(0,0,1,1)} = \dfrac{C(0,1,0,0) + C(0,1,0,1)}{C(0,1,1,0) + C(0,1,1,1)}$

(5b) $\quad \dfrac{C(1,0,0,0) + C(1,0,0,1)}{C(1,0,1,0) + C(1,0,1,1)} = \dfrac{C(1,1,0,0) + C(1,1,0,1)}{C(1,1,1,0) + C(1,1,1,1)}$

(5c) $\quad \dfrac{C(0,0,0,0) + C(0,0,1,0)}{C(0,0,0,1) + C(0,0,1,1)} = \dfrac{C(1,0,0,0) + C(1,0,1,0)}{C(1,0,0,1) + C(1,0,1,1)}$

(5d) $$\frac{C(0,1,0,0)+C(0,1,1,0)}{C(0,1,0,1)+C(0,1,1,1)} = \frac{C(1,1,0,0)+C(1,1,1,0)}{C(1,1,0,1)+C(1,1,1,1)}$$

One can substitute Equation (4b) for each instance of $C(s_A, s_B, r_A, r_B)$ in (5a-d), do some messy but elementary algebra, and solve for values of certain ratios of the detection probabilities $\lambda_A(s_A, r_A)$ and $\lambda_B(s_B, r_B)$:

(6) $\frac{\lambda_A(0,1)}{\lambda_A(0,0)} = 0.87409$, $\frac{\lambda_A(1,1)}{\lambda_A(1,0)} = 1.05054$, $\frac{\lambda_B(0,1)}{\lambda_B(0,0)} = 0.98185$, $\frac{\lambda_B(1,1)}{\lambda_B(1,0)} = 1.02008$

Applying any $\lambda_A(s_A, r_A)$ and $\lambda_B(s_B, r_B)$ that satisfy these ratios to $W_T(s_A, s_B, r_A, r_B)$ yields a set of counts $C(s_A, s_B, r_A, r_B)$ that correspond to the probabilities in Table 7. It is easy to check that these probabilities satisfy the no-signaling conditions. But the marginals $p(r_A = 0 | s_A = 0)$, $p(r_A = 0 | s_A = 1)$, $p(r_B = 0 | s_B = 0)$, and $p(r_B = 0 | s_B = 1)$ are not equal to 0.5, suggesting that the photon pairs are not maximally entangled.

Table 7: Probabilities Corresponding to Counts $C(s_A, s_B, r_A, r_B)$

| $s_A/r_A$ | $s_B/r_B$ | 0/0 | 0/1 | 1/0 | 1/1 |
|---|---|---|---|---|---|
| 0/0 | | 0.06475 | 0.42392 | 0.09073 | 0.39794 |
| 0/1 | | 0.42389 | 0.08744 | 0.41198 | 0.09935 |
| 1/0 | | 0.41756 | 0.07739 | 0.07376 | 0.42119 |
| 1/1 | | 0.07108 | 0.43397 | 0.42895 | 0.07610 |

I cannot guarantee that for every possible EPR experiment, a set of detection probabilities exists that will bring the observed counts into compliance with the no-signaling conditions, nor can I guarantee that if there are such detection probabilities, their ratios are unique. But for this experiment the ratios (6) are the unique ratios that accomplish the task.[5]

No simple necessary and sufficient conditions are known at present for whether the probabilities in Table 7 can be realized by quantum theory (Avis et al, 2008). However, Masanes (2003) provides necessary and sufficient conditions on the correlations (defined earlier as the probability Alice's and Bob's results agree less the probability they disagree, for each combination of settings $s_A$ and $s_B$), and the correlations from Table 7 do meet Masanes' conditions. The probabilities also meet conditions (41) in Navascues et al (2008), which are necessary but not sufficient for the probabilities (and not just the correlations) to have a quantum realization.

In light of the above results it may occur to the reader to wonder how Adenier & Khrennikov could conclude that the fair sampling assumption "cannot be maintained to be a reasonable assumption as it leads to an apparent violation of the no-signaling principle." In their normalization procedure, Adenier & Khrennikov assumed the one-sided versions of Equations (4a)—i.e., that the singles counts $D_A(s_A, r_A)$ and $D_B(s_B, r_B)$ (see Table 6) were fair samples from the counts that would have been obtained if the detectors were perfect, and if nothing triggered singles detections other than photons paired at the source. This assumption implies that for each $s_A$ and $s_B$:

(7) $\quad \dfrac{\lambda_A(s_A,1)}{\lambda_A(s_A,0)} = \dfrac{D_A(s_A,1)}{D_A(s_A,0)}, \quad \dfrac{\lambda_B(s_B,1)}{\lambda_B(s_B,0)} = \dfrac{D_B(s_B,1)}{D_B(s_B,0)}$

---

[5] I have performed similar calculations on the data for 30 other WJSWZ experiments, and there are unique ratios for each of them.

It is easy to check that conditions (7) are violated by any detection probabilities satisfying conditions (6) and the singles counts given in Table 6. By giving up conditions (7)—not an exorbitant cost, as I see it—one can have both no-signaling and fair-sampling.

## A LOCAL DETECTION DELAY MODEL

I have constructed a simple model that uses detection delays to explain how the wide and narrow windows can have different distributions of coincidences over the sixteen combinations of settings and results. I suppose that the time from the generation of a photon pair to Alice's (Bob's) detection of her (his) photon is random. To make the model local, I assume Alice's and Bob's delays are independent, and that Alice's (Bob's) delay time distribution depends only on her (his) setting and result. Define:

$g_A(t|s_A, r_A)$ = Distribution of time from generation of a photon pair to Alice's detection of her photon, given her setting and result $s_A, r_A$

$g_B(t|s_B, r_b)$ = Distribution of time from generation of a photon pair to Bob's detection of his photon, given his setting and result $s_B, r_B$

Of course, the time from the generation of a photon pair to an observer's detection of his photon must be positive, so $g_A(t|s_A, r_A) = g_B(t|s_B, r_B) = 0$ for all $t \leq 0$. Since they are probability distributions, they integrate to 1. Then according to this model:

(8) $\quad f(\Delta t|s_A, s_B, r_A, r_B) = \displaystyle\int_{-\infty}^{\infty} g_A(t|s_A, r_A) \times g_B(t + \Delta t|s_B, r_B) \times dt$

where $f(\Delta t|s_A, s_B, r_A, r_B)$ is the distribution of the time $\Delta t = t_B - t_A$ from Alice's to Bob's detection of photons in a true coincidence. Note that $\Delta t$ can be negative as well as positive. The estimated density of identified coincidences, $Cest(\Delta t|s_A, s_B, r_A, r_B)$, becomes:

(9) $\quad Cest(\Delta t|s_A, s_B, r_A, r_B) = W_T(s_A, s_B, r_A, r_B) \times f(\Delta t|s_A, s_B, r_A, r_B) + W_F(s_A, s_B, r_A, r_B)$

The use of $W_T(s_A, s_B, r_A, r_B)$ and $W_F(s_A, s_B, r_A, r_B)$ in Equation (9) ensures that the estimated counts in the wide window (i.e., integral of $Cest(\Delta t|s_A, s_B, r_A, r_B)$ over the wide window) must equal the raw counts from Table 3.

I choose Alice's and Bob's detection delay distributions so that the estimated density of coincidences is close to the observed density, $Cobs(\Delta t|s_A, s_B, r_A, r_B)$. That is, I find delay time distributions that make the error small in Equation (10).

(10) $\quad Cest(\Delta t|s_A, s_B, r_A, r_B) = Cobs(\Delta t|s_A, s_B, r_A, r_B) + Error(\Delta t|s_A, s_B, r_A, r_B)$

I solve a discretized version of this problem iteratively. I assume distributions $g_A(t|s_A, r_A)$ for Alice and solve a linear program to find the distributions $g_B(t|s_B, r_b)$ for Bob that minimize the sum of the absolute errors. I then hold Bob's distributions fixed and solve for a new set of distributions for Alice. I iterate until the distributions converge. There is no guarantee that this algorithm will converge in general, but it seems to converge if I start with distributions for Alice that consist of a single very narrow peak.

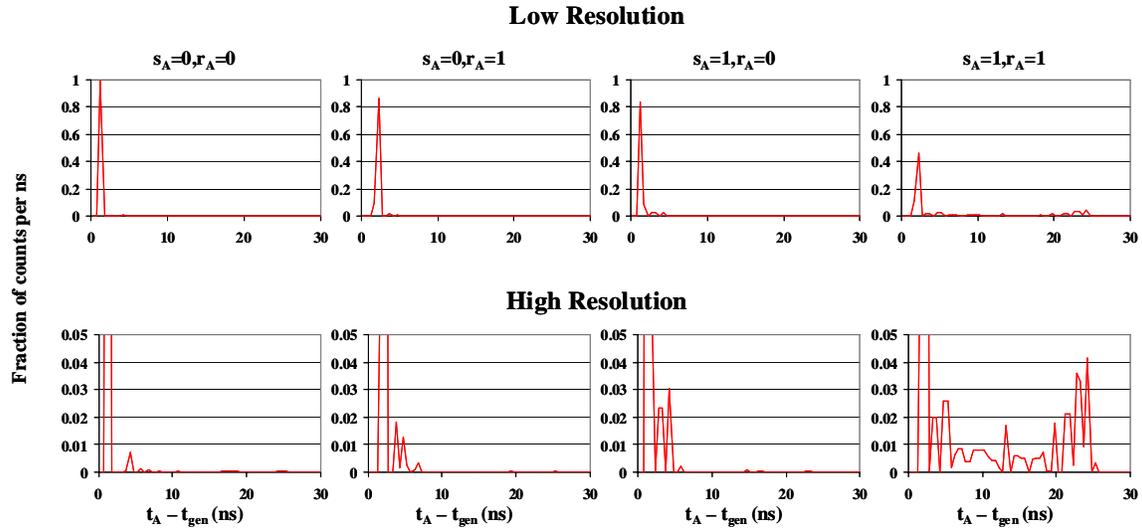

**Figure 3: Alice's Detection Delay Distributions**

Figures 3 and 4 show that the resulting delay distributions are mostly concentrated in a narrow peak. Expanding the vertical scale reveals that Alice's (Bob's) delay times have significant probabilities of deviating from the peak when $s_A = 1, r_A = 1$ ($s_B = 1, r_B = 1$). I can force my algorithm to generate smoother distributions at the cost of larger errors. But I have no reason save aesthetics to demand smoother distributions.

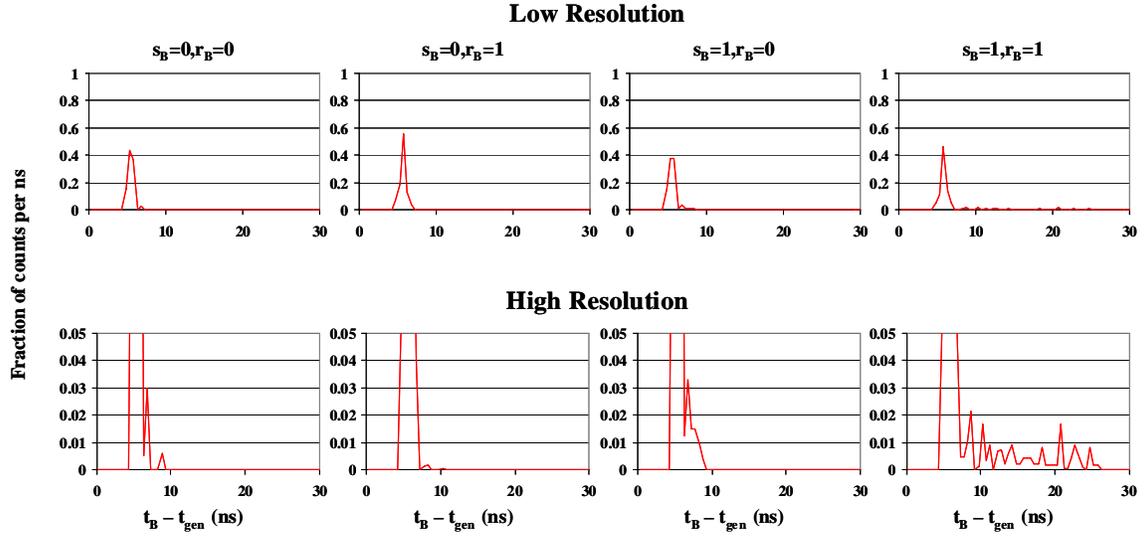

**Figure 4: Bob's Detection Delay Distributions**

These delay distributions produce pretty good approximations of the observed distributions of coincidences over the detection interval $\Delta t = t_B - t_A$ and the sixteen combinations of settings and results (Figures 5 and 6). Figure 6 shows the same data as Figure 5, with the vertical axis expanded.

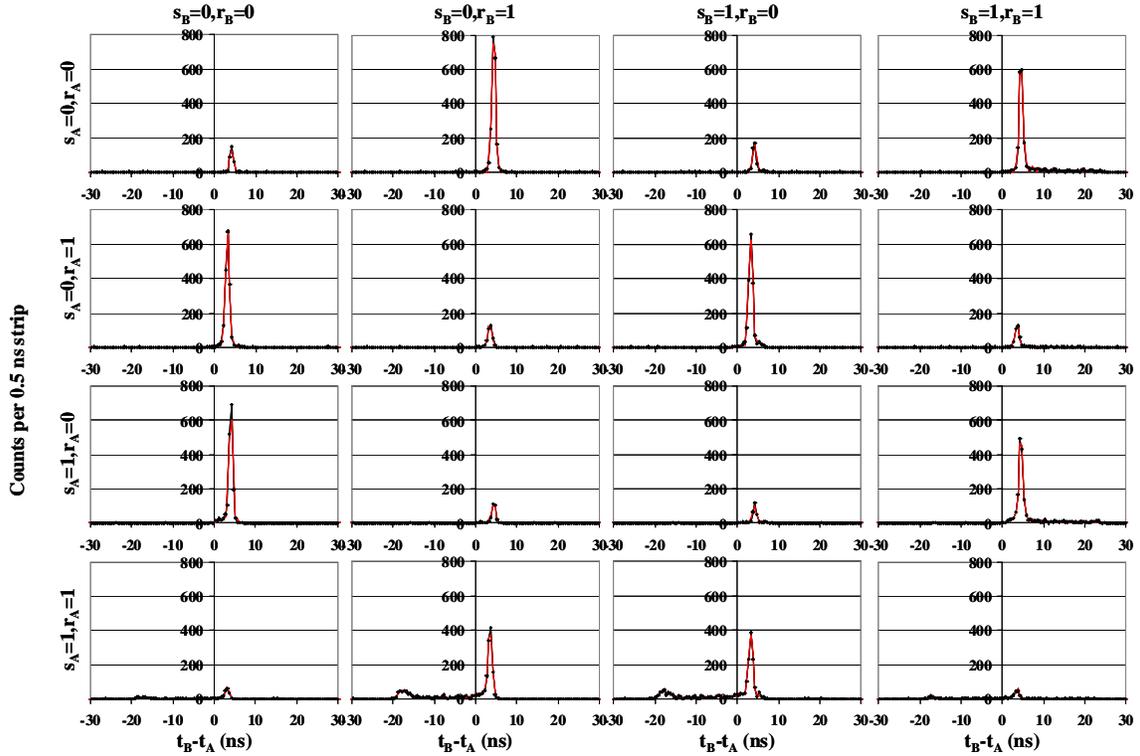

**Figure 5: Actual (black) vs. Calculated (red) Distributions of Coincidences, Low Resolution**

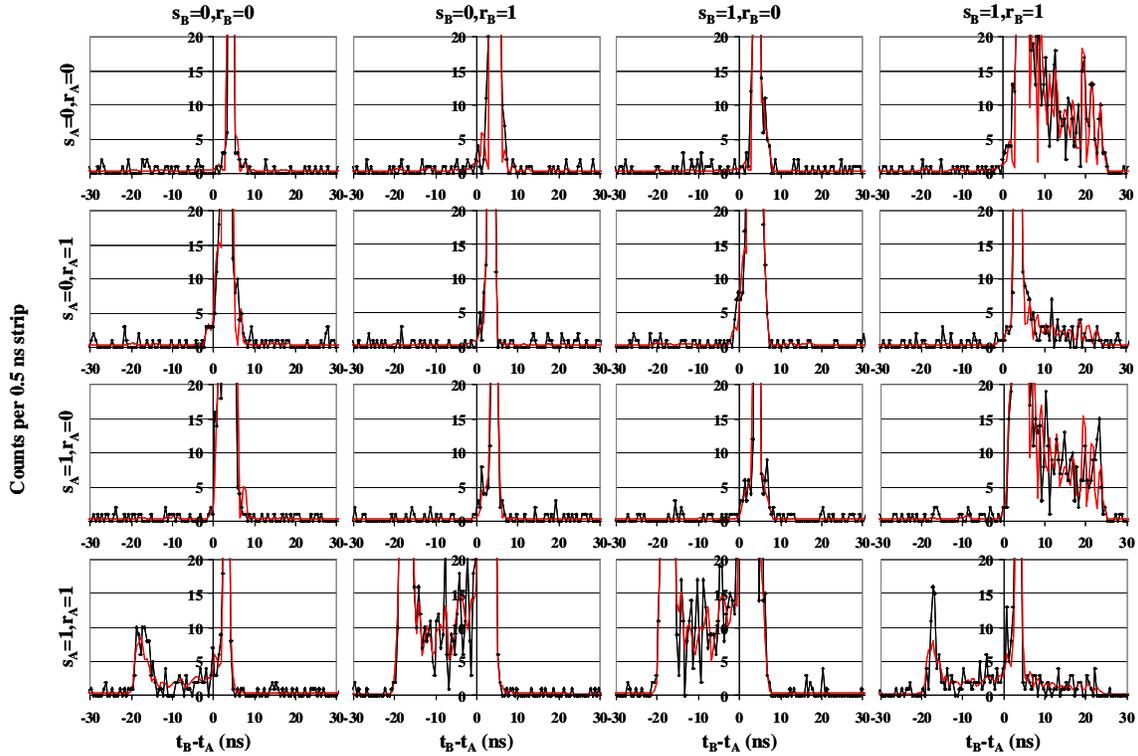

**Figure 6: Actual (black) vs. Calculated (red) Distributions of Coincidences, High Resolution**

## SUMMING UP

Weihs et al decided to register individual detections by Alice and Bob independently, and to analyze the data only after measurements were completed. In my view this decision was wise. The data can be analyzed and re-analyzed to test ideas not considered at the time of the experiment.

One such idea is that the coincidences one identifies depend on the detection window one selects. If the number of coincidences were merely scaled up and down as a function of the window width, this idea would not be very interesting. But it happens that the distribution of coincidences over settings and results changes as a function of the detection window.

This is the essence of the coincidence time loophole. In the literature, the loophole is generally used to explain how nature could post-select a distribution of coincidences that violates a Bell inequality from one that exhibits only classical correlations. But I did not find that wider, more inclusive windows violated Bell inequalities any less than narrow windows. So, while nature appears to have a mechanism to vary the difference in Alice's and Bob's detection times as a function of Alice's and Bob's settings and results, nature did not use the mechanism—at least, not in the WJSWZ experiments—to make a classical phenomenon simulate a quantum outcome.

But what is this mechanism?  The data are consistent with a purely local mechanism, one which varies the Alice's (Bob's) detection time of a photon solely as a function of Alice's (Bob's) setting and result, and for which the effects at Alice's and Bob's observation stations are independent.  Physically it could be due to differences in time measurement devices and detectors at each observation station.  There is no need to postulate that the photons of a pair carry any common information from the generation event to the detection events.

Disconcertingly, the coincidences identified by both the narrow and wide windows violate the no-signaling condition principle, though the violations are smaller for the wide window.  In 17 of the 31 experiments I analyzed, the wide window counts violate at least one of the four no-signaling conditions by at least 3 standard deviations.  Using a Normal approximation, there is a 1-tail probability of 0.001 that a violation that large would occur by chance.  Another six experiments show a violation of at least 2 standard deviations (probability 0.023), and each of the the remaining eight experiments shows a violation at least 1.17 standard deviations (probability of 0.121).  I have shown how detection inefficiencies could account for this, but I would rather find the missing coincidences.

It is conceivable that these coincidences might have been discarded.  Weihs (2007) reports that settings were changed randomly every 100 ns, and that detections that occurred during each 14 ns switching period were suppressed.  If these experiments are repeated, it might be worthwhile—and certainly not difficult—to collect these detections and mark them as occurring during a switching period.